\documentclass[prl,twocolumn,amsmath,amssymb,letterpaper,superscriptaddress]{revtex4} 
\usepackage{graphicx}
\usepackage{amsmath}
\usepackage{amsfonts}
\usepackage{amssymb}
\usepackage[colorlinks,linkcolor=blue,filecolor=blue,urlcolor=blue,citecolor=blue]{hyperref}

\begin{document}

\title{Normal-Superfluid Phase Separation in Spin-Half Bosons at Finite Temperature}

\author{Li He}
\affiliation{College of Physics and Electronic Engineering, Shanxi University, Taiyuan 030006, China}
\author{Peipei Gao}
\affiliation{Institute of Theoretical Physics, Shanxi University, Taiyuan 030006, China}
\affiliation{State Key Laboratory of Quantum Optics and Quantum Optics Devices, Shanxi University, Taiyuan 030006, China}
\author{Zeng-Qiang Yu}
\email{zqyu.physics@outlook.com}
\affiliation{Institute of Theoretical Physics, Shanxi University, Taiyuan 030006, China}
\affiliation{State Key Laboratory of Quantum Optics and Quantum Optics Devices, Shanxi University, Taiyuan 030006, China}

\begin{abstract}
For pseudospin-half bosons with inter-spin attraction and intra-spin repulsion, normal phase and  Bose condensed phase can coexist at finite temperature. The homogeneous system is unstable against the spinodal decomposition within a medium density interval, and consequently, a normal-superfluid phase separation takes place. The isothermal equation-of-state shows a characteristic plateau in the $P$-$V$ (pressure-volume) diagram, which is reminiscent of a classical gas-liquid transition, although, unlike the latter, the coexistence lines never terminate at a critical point as temperature increases. In a harmonic trap, the phase separation can be revealed by the density profile of the atomic cloud, which exhibits a sudden jump across the phase boundary.
\end{abstract}

\maketitle

{\it Introduction}.\,---
The relation between Bose-Einstein condensation (BEC) and gas-liquid condensation had been discussed for a long time~\cite{Uhlenbeck1938}. In Einstein's seminal work, BEC was thought as a spatial separation that the condensate neither occupy any volume nor contribute to pressure~\cite{Einstein1925}. Although the equation-of-state (EoS) of free bosons resembles that of a van~der~Waals gas in the transition region~\cite{Uhlenbeck1938,HuangBook}, such similarity is merely a coincidence due to the absence of interparticle interactions. As first pointed out by London, the condensation stemming from the Bose statistics is more appropriately understood in momentum space rather than in coordinate space~\cite{London1938}. Now, it is well known that a BEC transition associates the emergence of off-diagonal long range order~\cite{LevSandroBook}, while the classical gas-liquid condensation gives rise to the change only in density~\cite{LandauBook}.

Despite their distinctions in nature, the two kinds of condensation are not incompatible. In 1960, Huang proposed that, for bosons with hard-core repulsion and long range attraction, BEC transition and gas-liquid transition can take place simultaneously~\cite{Huang1960}. Based on an unrealistic model, this theory qualitatively reproduced the phase diagram of  $^4\text{He}$, in which the liquid phases, either superfluid or not, can coexist with the gas phase at finite temperature~\cite{LondonBook,He4Data}.

The recent realization of self-bound liquids with ultracold atoms opens up new perspective to explore the gas-liquid transition in the quantum degenerate regime~\cite{DipolarExp1,DipolarExp2,DipolarExp3,DipolarExp4,DropletExp1, DropletExp2,DropletExp3,DropletExp4}. Such liquids, which would collapse from the mean-field viewpoint, are stabilized by the many-body effects of quantum fluctuations~\cite{DropletTheory1}. At a balanced density, energy per particle reaches the minimum, and the liquid exhibits the unique self-bound character. Owing to the finite-size effect, a liquid drop is stable against evaporation only when its atom number exceeds a critical value. The bound-unbound transition has been experimentally observed in the dipolar condensates~\cite{DipolarExp3}, as well as the binary Bose mixtures~\cite{DropletExp1,DropletExp3,DropletExp4}.

\begin{figure}[ht!]
\includegraphics{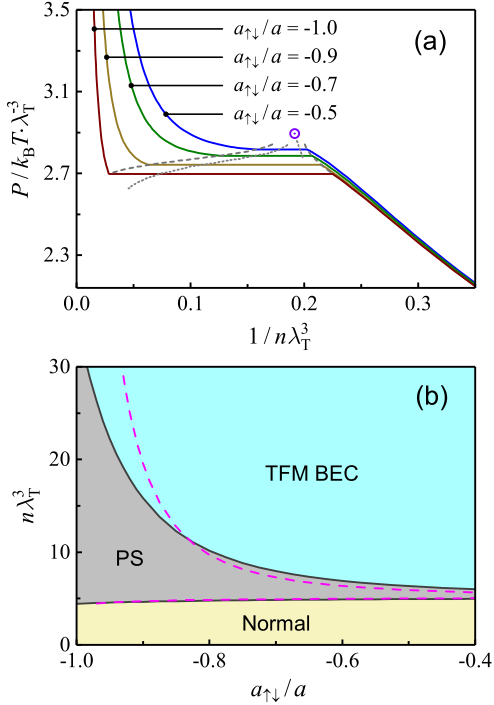}
\caption{(a) Isothermal EoS of spin-half bosons. The horizontal segments correspond to mixed states of the normal and the transverse ferromagnetic BEC phase. Coexistence lines (short-dashed) and spinodal lines (short-dotted) are plotted for $-1\leqslant a_{\uparrow\downarrow}/a \leqslant -0.3$. They are supposed to meet at a tricritical point (denoted by $\odot$) when the inter-spin attraction vanishes. (b) Phase diagram in terms of phase space density $n\lambda_{\rm T}^3$ and interaction parameter $a_{\uparrow\downarrow}/a$. Here, EoS and phase diagram are calculated based on the Popov theory; for comparison, phase boundaries predicted by the Hartree-Fock theory are also shown [dashed lines in (b)]. Parameters: $T=400$nK; for all the figures of this Letter, the mass of boson takes the value of $^{39}$K atom, and the intra-spin scattering length $a=65a_0$ (with $a_0$ the Bohr radius).} \label{Fig1}
\end{figure}

Previously, the liquid-like properties of two-component bosons have been investigated by many theoretical works. Most of them focus on the influence of quantum fluctuations at zero temperature~\cite{DropletTheory1,DropletTheory2, DropletTheory3,DropletTheory4,DropletTheory5, DropletTheory6,DropletTheory7,DropletTheory8,DropletTheory9}; few pay attention to thermal effects~\cite{DropletTheory7}. In this Letter, we study the thermodynamics of spin-half bosons with inter-spin attraction and intra-spin repulsion at finite temperature. Our main results are summarized in Fig.\;\ref{Fig1}. Reminiscent of classical gas-liquid condensation, the isotherms exhibit characteristic plateaus in the $P$-$V$ diagram, and the normal and the BEC phase coexist in the transition region. This unusual normal-superfluid phase separation (PS) is mainly driven by thermal fluctuations, although, quantum fluctuations still play an important role when the attractive and the repulsive mean-field energy almost cancel out. Contrary to the case of spinless bosons, here the BEC transition is of first order and is accompanied by an abrupt change in density; in this sense, the condensation occurs not only in momentum space but also in coordinate space.

\vspace{1ex}

{\it EoS of a Homogeneous System}.\,---
We consider weakly interacting bosonic atoms, which occupy two hyperfine sublevels labeled by the pseudospin $\sigma=\uparrow,\downarrow$. The grand-canonical Hamiltonian reads
\begin{align}
  \hat H = \sum_{\sigma} \int \mathrm{d}\mathbf{r} \, \hat\psi_\sigma^\dagger\left[ - \frac{\hbar^2\nabla^2}{2m} - \mu_\sigma + \sum_{\sigma'} \frac{g_{\sigma\sigma'}}{2} \hat\psi_{\sigma'}^\dagger \hat\psi_{\sigma'} \right] \hat\psi_\sigma \, ,
\end{align}
where $\hat\psi_\sigma$ and $\mu_\sigma$ are the field operator and the chemical potential, respectively, of $\sigma$-component, $m$ is the atomic mass, and $g_{\sigma\sigma'}$ are the interaction parameters in different spin channels. Within the Born approximation, $g_{\sigma\sigma'}=4\pi\hbar^2a_{\sigma\sigma'}/m$, with $a_{\sigma\sigma'}$ the corresponding scattering length. In the present work, we assume $\mu_\uparrow=\mu_\downarrow = \mu$ and $g_{\uparrow\uparrow}=g_{\downarrow\downarrow}=g$ (accordingly $a_{\uparrow\uparrow}=a_{\downarrow\downarrow}=a$). Under these considerations, the Hamiltonian possesses $\textrm{U}(1)\times\textrm{Z}_2$ symmetry in spin space, and the spontaneous magnetization would not emerge unless the symmetry is broken. Hereafter, we will focus on the regime of $a\geqslant-a_{\uparrow\downarrow}>0$, where the normal-superfluid PS could take place.

At finite temperature $T$, the importance of quantum degeneracy of a homogeneous system is governed by the phase space density $n\lambda_{\rm T}^3$, with $n$ the atomic density and $\lambda_{\rm T}=\sqrt{2\pi\hbar^2/mk_{\rm B}T}$ the thermal wavelength. For $n\lambda_{\rm T}^3\ll 1$, the quantum effects are negligible, and atoms behave like classical particles with the EoS given by
\begin{equation}
  P = nk_{\rm B}T + \tfrac{1}{4} \left(2g+g_{\uparrow\downarrow}\right)n^2 ,  \label{EoS_low_n_limit}
\end{equation}
where the first term coincides with the pressure of an ideal classical gas, and the second term corresponds to the leading order contribution of interactions. On the other hand, for $n\lambda_{\rm T}^3 \gg 1$, the system is in the highly degenerate regime, where almost all the bosons condense in the zero-momentum state~\cite{footnote_high_density}. The spinor condensate breaks the rotation symmetry about the $\sigma_z$-axis with the wavefunction given by
$\varphi = \sqrt{n/2} \left(e^{-i\theta_\uparrow},e^{-i\theta_\downarrow}\right)^{\dagger}$.
Within the Bogoliubov approximation,  we obtain the EoS
\begin{equation}
  P = \frac{1}{2}g_+n^2 + \frac{32n^2}{5\sqrt{\pi}} \Big(  g_+ \sqrt{na_+^3} + g_- \sqrt{na_-^3} \Big) \, , \label{EoS_high_n_limit}
\end{equation}
with $g_\pm=\frac{1}{2}(g\pm g_{\uparrow\downarrow})$ and $a_\pm=\frac{1}{2}(a\pm a_{\uparrow\downarrow})$. The first term of (\ref{EoS_high_n_limit}) is the mean-field energy of the condensate, and the second term is the Lee-Huang-Yang correction originating from the quantum fluctuations.

For a constant $T$, as density varies from small to large, the EoS will change its form from (\ref{EoS_low_n_limit}) to (\ref{EoS_high_n_limit}). During this evolution, two phase transitions take place successively: one is the transverse ferromagnetic (TFM) transition at critical density $n_{\rm M}$, the other is the BEC transition at critical density $n_{\rm C}$. The transverse spin polarization is energetically favorable owing to the inter-spin attraction, and the resulting phase transition occurs for arbitrary small negative $a_{\uparrow\downarrow}$~\cite{Ashhab2005,Radic2014}. At the mean-field level, the onset of ferromagnetism is fixed by the condition~\cite{note_supplementary}
\begin{align}
  2 a_{\uparrow\downarrow} {\rm Li}_{1/2}\big(e^{\beta\mu'}\big) + \lambda_{\rm T}= 0 \, ,
  \label{magnetic_transition}
\end{align}
where $\text{Li}_q (z)=\sum_{\ell=1}^\infty{z^\ell/\ell^q}$
is the polylogarithm function of $q$~order, $\beta=1/k_{\rm B}T$, and $\mu'=\mu-(g+\frac{1}{2}g_{\uparrow\downarrow})n$. By combining Eq.~(\ref{magnetic_transition}) with the density equation $n\lambda_{\rm T}^3 = 2\,\mathrm{Li}_{3/2}\big(e^{\beta\mu'}\big) $, the critical point $n_{\rm M}$ thus can be determined. Up to the first order of $a_{\uparrow\downarrow}$, we find
\begin{equation}
  n_{\rm M} = n_{\rm C}^{(0)} + 8\pi a_{\uparrow\downarrow}/\lambda_\mathrm{T}^4 \, ,
\end{equation}
where $n_\mathrm{C}^{(0)} = 2 \zeta (\tfrac{3}{2}) \lambda_\mathrm{T}^{-3} \simeq 5.22 \lambda_\mathrm{T}^{-3}$ is the critical density of the BEC transition in the noninteracting case, with $\zeta(q)$ the Riemann zeta function.

\begin{figure}
\includegraphics{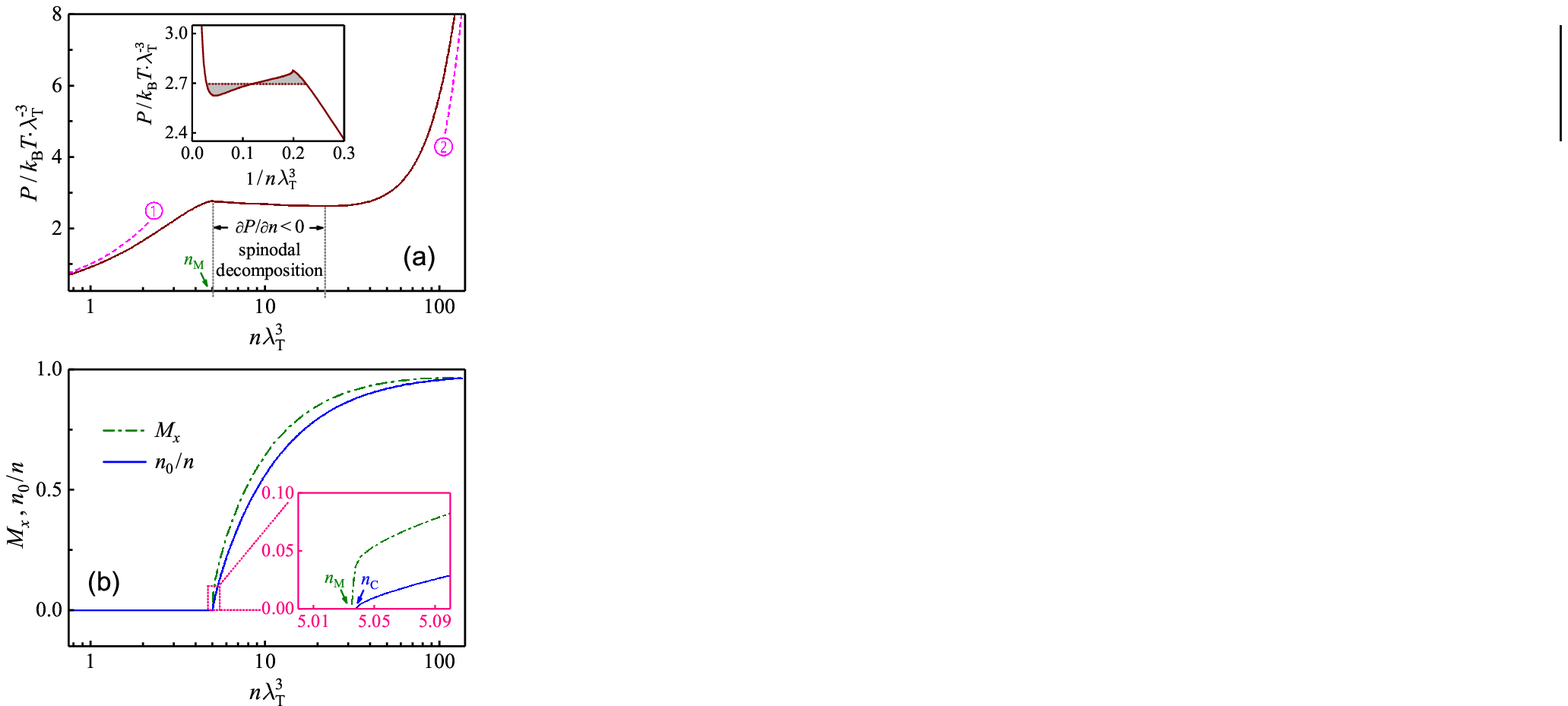}
\caption{(a) Isothermal $P$-$n$ curve of a  hypothetically homogeneous system predicted by the Popov theory.
Short-dashed lines labelled by \raisebox{.5pt}{\textcircled{\raisebox{-.5pt} {\footnotesize 1}}} and \raisebox{.5pt}{\textcircled{\raisebox{-.5pt} {\footnotesize 2}}} correspond to the asymptotic EoS (\ref{EoS_low_n_limit}) and (\ref{EoS_high_n_limit}), respectively. Inset: the Maxwell construction in the coexistence region.  (b) Transverse magnetization $M_x$ and condensate fraction $n_0/n$ as functions of $n\lambda_{\rm T}^3$. Inset shows the critical points $n_\mathrm{M}$ and $n_\mathrm{C}$ are very close. Parameters: $a_{\uparrow\downarrow}=-a$, and $T=400$nK.} \label{Fig2}
\end{figure}

Bose condensation emerges when density exceeds the higher threshold value $n_{\rm C}$.
For weak inter-spin attraction, $n_{\rm C}$ is very close to $n_{\rm M}$, and the leading difference between them is of order $a_{\uparrow\downarrow}^2/\lambda_{\rm T}^5$~\cite{note_supplementary}. In the BEC phase, both thermal atoms and condensate contribute to the transverse magnetization, and their spin polarization prefer to align in the same direction~\cite{Ashhab2005}. Without any loss of generality, we set the polarization along the $\sigma_x$-axis. The magnetization $M_x$ and the condensate fraction $n_0/n$, which act as the order parameters associating with the respective phase transitions, can be obtained from the Popov theory or the Hartree-Fock (HF) theory~\cite{note_supplementary}. Their rises with $n\lambda_{\rm T}^3$ are displayed in Fig.\;\hyperref[Fig2]{2(b)}.

From EoS (\ref{EoS_low_n_limit}) and (\ref{EoS_high_n_limit}), it is easy to show that the isothermal compressibility $\kappa_T=(\partial n/\partial P)_T/n$ is positive in both cases of $n\lambda_{\rm T}^3\ll 1$ and $n\lambda_{\rm T}^3\gg 1$, hence the system is mechanically stable under the corresponding densities. However, such stability no longer survives when density is within a medium range~\cite{note_instability}. The onset of the spinodal decomposition occurs at $n_{\rm M}$, where $\kappa_T$ exhibits a discontinuity. For $|a_{\uparrow\downarrow}|\ll\lambda_{\rm T}$, we find~\cite{note_supplementary}
\begin{align}
 \left. \kappa_T^{-1}\right|_{n\rightarrow n_{\rm M}^-} &=  gn_{\rm M}^2 \, , \\
 \left. \kappa_T^{-1}\right|_{n\rightarrow n_{\rm M}^+} &=  \left(g + \tfrac{1}{2} g_{\uparrow\downarrow} -  C \lambda_{\rm T}^3  k_{\rm B}T \right)n_\mathrm{M}^2 \, ,  \label{compressibility_nM}
\end{align}
with $C=-1/2\zeta\big(\tfrac{1}{2}\big)\simeq 0.34$. In the weakly interacting regime, the last term in the parentheses of Eq.\;(\ref{compressibility_nM}) is dominant, therefore the sign of $\kappa_T$ changes across the TFM transition. For $|a_{\uparrow\downarrow}|$ and $a$ being comparable, compressibility retains a negative value within a wide range of density extending from the normal phase to the BEC phase [see Fig.\;\hyperref[Fig2]{2(a)}].

\vspace{1ex}

\textit{Normal-Superfluid PS}.\,---
The mechanical instability discussed above implies a PS between the normal and the ferromagnetic BEC phase at finite temperature. The coexistence region can be fixed by the Maxwell construction~\cite{HuangBook,LandauBook}, with an example illustrated in the inset of Fig.\;\hyperref[Fig2]{2(a)}.
Alternatively, one can determined the transition line according to the balanced conditions for chemical potential and pressure
\begin{equation}\label{ps_condition}
  \mu(n_1,T)=\mu(n_2,T) \, , \qquad
  P(n_1,T)=P(n_2,T) \, ,
\end{equation}
where $n_1$ and $n_2$ are densities of the normal and the BEC phase, respectively, in the mixed state ($n_1<n_{\rm M}$, $n_2>n_{\rm C}$). For average density within the interval $n_1<n<n_2$, the mixed state has a lower free energy than that of the homogeneous ones, and as a result, the normal-superfluid PS takes place. Remarkably, the isotherm shows a horizontal segment in the coexistence region, which is usually recognized as a characteristic of gas-liquid condensation [see Fig.\;\hyperref[Fig1]{1(a)}]. Using the thermodynamic relation $\mathrm{d}P = s\mathrm{d}T + n\mathrm{d}\mu$ ($s$ is the entropy density),  one can verify that the coexistence pressure satisfies the Clapeyron equation $ \mathrm{d}P/\mathrm{d}T = L/T\big(n_1^{-1}-n_2^{-1}\big)$, with $L$ the latent heat per particle of the phase transition~\cite{HuangBook,LandauBook}.

When the inter-spin attraction is tuned from weak to strong, the PS region extends to an increasingly large area in the phase diagram. For $|a_{\uparrow\downarrow}|<0.9a$, the results predicted by the Popov theory and the HF theory are essentially the same, which suggests the PS in this regime is mainly driven by thermal fluctuations [see Fig.\;\hyperref[Fig1]{1(b)}]. For stronger inter-spin attraction, however, quantum fluctuations play an important role to affect the phase boundary. In particular, at $a_{\uparrow\downarrow}=-a$, where the mean-field energy of condensate totally vanishes, the HF theory wrongly predicts the BEC phase would always collapse at finite temperature. This deficiency can be remedied by the Popov theory, in which the Lee-Huang-Yang correction due to the quantum fluctuations is properly taken into account~\cite{note_supplementary}.

\begin{figure}
\includegraphics{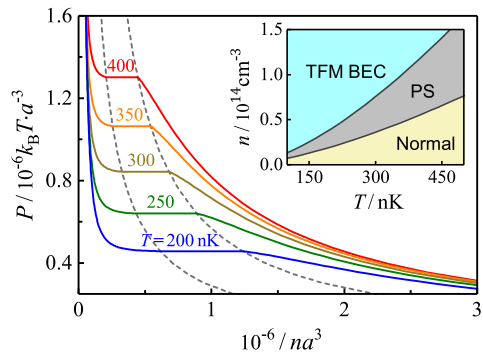}
\caption{Isotherms at different temperatures for $a_{\uparrow\downarrow}=-0.8a$. Coexistence lines (short-dashed) never terminate as $T$ increases. Inset: phase diagram in the $T$-$n$ plane. For parameters used here, the results predicted by the Popov theory and the Hartree-Fock theory are almost identical.} \label{Fig3}
\end{figure}

For fixed $a$ and $a_{\uparrow\downarrow}$, the PS occurs at any non-zero~$T$, and
the density difference between the coexisting normal and BEC phase becomes more pronounced as $T$ increases (see Fig.\;\ref{Fig3}). This feature is in bold contrast to the case of classical gas-liquid condensation, where gas and liquid become indistinguishable above a critical temperature.
Actually, the two separated phases considered here differ not only in density but also in symmetry. Across the phase interface, the order parameters $n_0/n$ and $M_x$ exhibit abrupt changes as well.
In this respect, the normal-superfluid PS is more like the gas-liquid coexistence in $^4$He below the lambda point, where the gas behaves almost classically, while the liquid (He-II) shows a unique quantum nature~\cite{LondonBook,He4Data}.

It is well known that mean-field approaches, such as the Popov theory and the HF theory, usually lead to a BEC transition of first order. The normal-superfluid PS was also predicted for spinless bosons with purely repulsive interactions~\cite{Huang1957,ReentranceMF1,ReentranceMF2}, in which case the false spinodal decomposition near the BEC transition is due to the unphysical multi-valued behavior of the mean-field EoS. Such an artifact differs from our situation, where the mechanical instability is attributed to the inter-spin attraction. As the attraction strength increases, the density interval of spinodal decomposition enlarges, while the artificial multi-valued region tends to vanish~\cite{note_supplementary}. For $|a_{\uparrow\downarrow}|$ and $a$ being comparable, the phase-separated densities $n_1$ and $n_2$ turn out to be considerably away from $n_{\rm C}$. In that case, the mean-field description is supposed to be qualitatively reliable. On the other hand, the critical fluctuations beyond the mean-field level will become crucial when the PS region shrinks to small. Intuitively, the BEC transition should be of second order as usual if the inter-spin interaction turns into repulsion. Thus, a tricritical point is expected to appear at $a_{\uparrow\downarrow}=0$~\cite{footnote_tricritical_point}. This interesting tricriticality can be revealed by more sophisticated methods~\cite{Holzmann2003,Prokofev2004}, and we leave the issue for future study.

\vspace{1ex}

\begin{figure}
\includegraphics{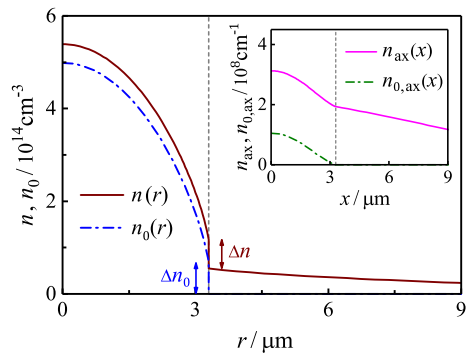}
\caption{Density and condensate density profiles of $N=5\times10^5$ atoms in an isotropic harmonic trap with trapping frequency $\omega=2\pi \times 150$Hz. The vertical short-dashed line indicates the phase boundary. Inset: integrated axial density profiles. Parameters: $a_{\uparrow\downarrow}=-0.8a$, and $T=400$nK.}  \label{Fig4}
\end{figure}

\textit{Density Profiles in a Trap}.\,---
Now, we discuss the experimental relevance of our theory. In a harmonic trap, the normal-superfluid PS can be readily observed through the density profile of the atomic cloud. Below the condensation temperature, the BEC phase, being of a relatively higher density, would occupy the central region of the trap and be surrounded by an outer rim of the normal phase. Such shell structure is illustrated in Fig.\;\ref{Fig4}, based on the mean-field calculation combined with the local density approximation (LDA). In contrast to the usual bimodal distribution, both the total and the condensate density profile exhibit a sudden jump at the phase boundary, which is a clear signature of the PS~\cite{footnote_density_jump}. Experimentally, the 3D density distribution can be achieved by means of \textit{in}-\textit{situ} absorption imaging followed by an inverse Abel transform.  This method has been previously employed to detect the PS in spin imbalanced Fermi gases~\cite{MIT2006,Rice2006,MIT2008,Rice2011}. We note that, in actual experiments, the phase interface would become less sharp due to the surface tension effects. Nevertheless, a steep change in density distribution would be still discernible for sufficiently large systems.

Within the LDA, the featured EoS can also be extracted from the density profiles~\cite{Yip2007,Ho2010,ENS2010-1,ENS2010-2,ENS2011}. Along certain axial direction, say, the $x$-axis, the local pressure can be obtained via the formula~\cite{Yip2007,Ho2010}
\begin{align}
  P(x,0,0) = m\omega_y\omega_z n_{\rm ax} (x) / 2\pi \, ,
\end{align}
where $\omega_i$ are the frequencies of the harmonic trap $(i=x,y,z)$, and $n_{\rm ax}(x)=\int \mathrm{d}y\mathrm{d}z\, n({\bf r})$ is the integrated axial density. The first-order nature of the normal-superfluid transition implies the discontinuity of the derivative of pressure, which results in a kink in the axial density profiles at the phase boundary (see the inset of Fig.\;\ref{Fig4}). This singular behavior can also be inferred from the relation $\mathrm{d}n_{\rm ax}(x)/\mathrm{d}x = -2\pi x n(x,0,0)\omega_x^2/\omega_y\omega_z$ \cite{Yip2007,Ho2010}, according to which, the abrupt change of the slope of $n_{\rm ax}(x)$ is proportional to the jump amplitude of the 3D density.

\vspace{1ex}

\textit{Discussion and Conclusion}.\,---
It should be noted that, in the transverse ferromagnetic BEC phase, atoms occupy superpositions of the hyperfine sublevels labeled by $\uparrow$ and $\downarrow$. From the viewpoint of symmetry breaking, the emergence of the TFM order requires a weak perturbation of spin-flip.  Experimentally, such circumstance can be realized when an applied radio-frequency field driving the inter-spin transition is adiabatically switched off, and the magnetization then can be measured by using a $\pi$/2 pulse, which rotates the transverse spin to the $\sigma_z$-direction~\cite{SpinExp1,SpinExp2,SpinExp3,SpinExp4}.

On the other hand, if the system is initially prepared without the inter-spin coupling, the transverse  ferromagnetism would not occur spontaneously, as the atom number in each hyperfine sublevel is individually conserved. The incoherent mixture achieved in this way also undergoes a normal-superfluid PS, although, the coexistence region shrinks with respect to the spin-half system, and the superfluid phase has a higher free energy than that of the ferromagnetic BEC~\cite{note_supplementary}. For heteronuclear mixtures, the situation is even more complicated, since the mass ratio of the constituent atoms may have additional effects on the mechanical and the diffusive stabilities. The possible occurrence of a similar PS in dual-species Bose mixtures, such as $^{41}$K-$^{87}$Rb~\cite{DropletExp4}, $^{39}$K-$^{87}$Rb~\cite{Mixture_Exp1}, and $^{23}$Na-$^{87}$Rb~\cite{Mixture_Exp3}, will be considered elsewhere.

We finally mention that the PS studied here differs from the immiscible phenomenon, which is governed by the same Hamiltonian, but with repulsive interspecies interactions~\cite{MiscibleExp1,MiscibleExp2,MiscibleExp3,MiscibleExp4, MiscibleTheory1,MiscibleTheory2,MiscibleTheory3,MiscibleTheory4,MiscibleTheory5, MiscibleTheory6,MiscibleTheory7,MiscibleTheory8,MiscibleTheory9,MiscibleTheory10}. In that case, the two separated phases are both superfluid, and across their interface, the longitudinal magnetization shows an abrupt change. The miscible-immiscible transition occurs when the positive $a_{\uparrow\downarrow}$ reaches a critical value. Recent investigation reveals this critical condition is significantly affected by the interaction driven thermal fluctuations~\cite{MiscibleTheory10}.

In summary, we have shown that, for spin-half bosons with both attractive and repulsive interactions, the normal phase can coexist with the superfluid BEC phase at finite temperature, and the isotherms exhibit the characteristics of gas-liquid condensation. Our predictions for the EoS and the phase diagram can be examined in current experiments with ultracold atoms. A further interesting issue concerns the extension of this study to the regime of $a_{\uparrow\downarrow}<-a<0$, where the spinodal decomposition occurs even at zero temperature. Presumably, as the density of the normal phase tends to vanish, the quantum liquid will eventually become self-bound in free space.

\vspace{1ex}

\acknowledgments{ \textit{Acknowledgments}.\,---
We thank Lan Yin, Wei Yi, Hui Zhai, Zhenhua Yu, Shizhong Zhang, and Jing Zhang for helpful discussions. This work is supported by NSFC under Grant No.~11674202, the Applied and Fundamental Research Program of Shanxi Province under Grants No.~201601D011014 and 201901D211187,  and the Fund for Shanxi 1331 KSC Project.}


\begin{thebibliography}{99}
\bibitem{footnote_high_density}
More precisely, the density considered here is within the range of $\lambda_{\rm T}^{-3} \ll n \ll (a\pm a_{\uparrow\downarrow})^{-3}$, thus the BEC state is still weakly interacting.
\bibitem{note_instability}
Here, we focus on the mechanical instability. The diffusive stability is always guaranteed, since the inter-spin attraction favors balanced spin populations.
\bibitem{footnote_tricritical_point}
In Fig.\;\hyperref[Fig1]{1(a)}, the pressure and the density of the tricritical point are estimated by the HF theory with $a_{\uparrow\downarrow}$ set to zero. The size of the symbol is greater than the uncertainty due to the multi-valued behavior of the mean-field EoS.
\bibitem{footnote_density_jump}
The density of thermal atoms also has a jump, but with a much smaller amplitude.
\end{thebibliography}
\end{document}


\centerline{\textbf{Supplementary Material for}}

\title{Normal-Superfluid Phase Separation in Spin-Half Bosons at Finite Temperature}

\author{Li He}
\affiliation{College of Physics and Electronic Engineering, Shanxi University, Taiyuan 030006, China}
\author{Peipei Gao}
\affiliation{Institute of Theoretical Physics, Shanxi University, Taiyuan 030006, China}
\affiliation{State Key Laboratory of Quantum Optics and Quantum Optics Devices, Shanxi University, Taiyuan 030006, China}
\author{Zeng-Qiang Yu}
\affiliation{Institute of Theoretical Physics, Shanxi University, Taiyuan 030006, China}
\affiliation{State Key Laboratory of Quantum Optics and Quantum Optics Devices, Shanxi University, Taiyuan 030006, China}

\maketitle

\renewcommand{\theequation}{S\arabic{equation}}
\renewcommand{\thefigure}{S\arabic{figure}}

\section{I.\quad T\lowercase{ransverse} F\lowercase{erromagnetic} P\lowercase{hase} T\lowercase{ransition}}

For the spin-half Bose system with inter-spin attractive interactions, as its phase space density $n\lambda_{\rm T}^3$ increases, the occurrence of transverse ferromagnetism precedes the BEC transition~\cite{Ashhab2005,Radic2014}. Within the Hartree-Fock approximation, the transverse magnetization in the non-condensed phase satisfies the self-consistency equation
\begin{align}
  n\lambda_{\rm T}^3 M_x &= {\rm Li}_{3/2}\big(e^{\beta\mu'-\beta g_{\uparrow\downarrow}nM_x/2}\big) - {\rm Li}_{3/2}\big(e^{\beta\mu'+\beta g_{\uparrow\downarrow}nM_x/2}\big) \, . \label{self-consist-0}
\end{align}
Here, without any loss of generality, we assume the spin polarization along the $\sigma_x$-axis. The magnetization $M_x$ and the shifted chemical potential $\mu'=\mu-(g+\frac{1}{2}g_{\uparrow\downarrow})n$ can be evaluated from Eq.\;(\ref{self-consist-0}) combined with the density equation
\begin{align}
  n\lambda_{\rm T}^3 &= {\rm Li}_{3/2}\big(e^{\beta\mu'-\beta g_{\uparrow\downarrow}nM_x/2}\big) + {\rm Li}_{3/2}\big(e^{\beta\mu'+\beta g_{\uparrow\downarrow}nM_x/2}\big) \, . \label{density-eqn-0}
\end{align}
Apparently, $M_x=0$ is always a possible solution of these equations, which corresponds to the non-magnetic state. Actually, this is the only solution for sufficiently small phase space density. When $n\lambda_{\rm T}^3$ exceeds a critical value, an additional solution with finite magnetization and lower free energy appears, which gives rise to the emergence of the ferromagnetic order.

For $|a_{\uparrow\downarrow}| \ll \lambda_{\rm T}$, the non-condensed ferromagnetic phase exists only in a narrow window of $n\lambda_{\rm T}^3$ with small magnetization. In that case, Eqs.\;(\ref{self-consist-0}) and (\ref{density-eqn-0}) can be solved through the series expansion in powers of $M_x$. Retaining the terms up to the third order, we find
\begin{align}
  \left(g_{\uparrow\downarrow}nM_x\right)^2 = 
  -24 \left[ g_{\uparrow\downarrow} \partial_{\mu'} {\rm Li}_{3/2} \big(e^{\beta \mu'}\big) + \lambda_{\rm T}^3  \right] / g_{\uparrow\downarrow} \partial_{\mu'}^3 {\rm Li}_{3/2} \big(e^{\beta \mu'}\big) \, ,  \label{Mx_solution}
\end{align}
with $\mu'$ determined by
\begin{align}
  n\lambda_{\rm T}^3 \,=\, 2\, {\rm Li}_{3/2} \big(e^{\beta \mu'}\big) - 6 \; \partial_{\mu'}^2 {\rm Li}_{3/2} \big(e^{\beta \mu'}\big) \left[ g_{\uparrow\downarrow} \partial_{\mu'} {\rm Li}_{3/2} \big(e^{\beta \mu'}\big) + \lambda_{\rm T}^3  \right]  / g_{\uparrow\downarrow} \partial_{\mu'}^3 {\rm Li}_{3/2} \big(e^{\beta \mu'}\big) \, .  \label{density-eqn-small-mx}
\end{align}
The critical condition for the ferromagnetic transition is then obtained by setting $M_x$ equal to zero in (\ref{Mx_solution}), i.\,e.,
\begin{align}
  g_{\uparrow\downarrow} \partial_{\mu'} {\rm Li}_{3/2} \big(e^{\beta \mu'}\big)  + \lambda_{\rm T}^3 = 0 \, . \label{FM-transition}
\end{align}
Using the identity $ \partial_{x} {\rm Li}_q(e^{x}) = {\rm Li}_{q-1}(e^{x}) $, we can rewrite the condition (\ref{FM-transition}) in the form of Eq.\;(4) of the main text.

With the function $n(\mu')$ derived in~(\ref{density-eqn-small-mx}), the isothermal compressibility $\kappa_T = \left(\partial n /\partial \mu \right)_T /n^2$ can be evaluated straightforward. At a given $T$, when density approaches the critical point $n_{\rm M}$ from above, we find
\begin{align}
  \left. \frac{\partial n}{\partial \mu'} \right|_{n\rightarrow n_{\rm M}^+} = \, \lambda_{\rm T}^{-3} \left\{ 2\partial_{\mu'} {\rm Li}_{3/2} \big(e^{\beta \mu'}\big) - 6 \big[ \partial_{\mu'}^2 {\rm Li}_{3/2} \big(e^{\beta \mu'}\big)\big]^2 / \partial_{\mu'}^3 {\rm Li}_{3/2} \big(e^{\beta \mu'}\big) \right\}_{n=n_{\rm M}}
 = \, \beta\lambda_{\rm T}^{-3}  \left[ 2\zeta\big(\tfrac{1}{2}\big) + \mathcal{O} \big(\beta\mu' \big) \right] \, ,
  \label{partial-n-mu}
\end{align}
where, in the second equality, we have used the fact that $\beta\mu'$ is negative small at the ferromagnetic transition, and accordingly, the polylogarithm function can be simplified as~\cite{Robinson1951}
\begin{align}
  {\rm Li}_{3/2} \left(e^{x}\right) = \zeta \big(\tfrac{3}{2} \big) - 2 \sqrt{\pi |x|} + \zeta\big(\tfrac{1}{2}\big)\, x + \mathcal{O} \left( x^2 \right) \, ,  \qquad (x\rightarrow 0^-) \, .
  \label{expansion}
\end{align}
By applying the expansion (\ref{expansion}) to the critical condition (\ref{FM-transition}), it is readily to show that the omitted terms in (\ref{partial-n-mu}) are at least at order of $a_{\uparrow\downarrow}^2/\lambda_{\rm T}^2$. Therefore, up to first order of the interaction parameters, the inverse compressibility can be written as Eq.~(7) of the main text, namely
\begin{align}
  \left. \kappa_T^{-1} \right|_{n\rightarrow n_{\rm M}^+} = \left( g+ \tfrac{1}{2}g_{\uparrow\downarrow} \right)n_{\rm M}^2 + n_{\rm M}^2 \lambda_{\rm T}^3/2\beta\zeta\big(\tfrac{1}{2}\big) \, . \label{inverse_kappa}
\end{align}
For $|a_{\uparrow\downarrow}|\ll \lambda_{\rm T}$, the last negative term on the r.h.s of (\ref{inverse_kappa}) is dominant, which results in the mechinical instability of the ferromagnetic phase\;\footnote{The collapse of the TFM phase was also predicted in Ref.~\cite{Radic2014}, but the normal-superfluid PS was not discussed there.}.

On the other hand, in the non-magnetic phase with $n<n_{\rm M}$, $\mu'$ satisfies the density equation $n\lambda_{\rm T}^3=2 {\rm Li}_{3/2} \big(e^{\beta \mu'}\big)$, which yields  $\partial n/ \partial \mu' = 2\beta\lambda_{\rm T}^{-3}{\rm Li}_{1/2} \big(e^{\beta \mu'}\big)>0$. Hence, the compressibility $\kappa_T$ exhibits a discontinuity with a sign change at the ferromagnetic phase transition.

%

\section{II.\quad P\lowercase{opov} T\lowercase{heory and} H\lowercase{artree}-F\lowercase{ock} T\lowercase{heory}}

We employ the Popov theory and the Hartree-Fock theory to study the thermodynamics of spin-half bosons at finite temperature. The formalism has been presented in a previous work by one of us~\cite{Yu2013}. In the transverse ferromagnetic BEC phase, we introduce the mean-field parameters $\delta n$ and $\delta M_x$ to denote the density and magnetization of thermal atoms, respectively. These quantities can be determined self-consistently from the equations
\begin{align}
  \delta n &=  \sum_{\alpha=\pm} \int \frac{d^3k}{(2\pi)^3} \left[ \left(u_{k,\alpha}^2 + v_{k,\alpha}^2\right) f (E_{k,\alpha}) + v_{k,\alpha}^2 \right] \, , \label{self-consist-1}\\
  \delta n\delta M_x &= \sum_{\alpha=\pm} \alpha \int \frac{d^3k}{(2\pi)^3} \left[ \left(u_{k,\alpha}^2 + v_{k,\alpha}^2 \right)  f (E_{k,\alpha}) + v_{k,\alpha}^2 \right] \, , \label{self-consist-2}
\end{align}
where $f(E_{k,\alpha})=1/(e^{\beta E_{k,\alpha}}-1)$ is the Bose distribution function with $E_{k,\alpha}$ being the excitation energy of quasi-particles. In the Popov theory, $ E_{k,+} = \sqrt{\epsilon_k \left(\epsilon_k+2g_+n_0\right)} $, $E_{k,-} = \sqrt{(\epsilon_k +g_-n_0 -g_{\uparrow\downarrow}\delta n \delta M_x)^2-g_-^2n_0^2}$,
$u_{k,\pm}$ and $v_{k,\pm}$ are coefficients of the Bogoliubov transformation satisfying the relations $u_{k,\pm}^2-v_{k,\pm}^2=1$ and $2u_{k,\pm}v_{k,\pm}=-g_\pm n_0/E_{k,\pm}$. In the Hartree-Fock theory, $E_{k,+}=\epsilon_k+g_+n_0$, $E_{k,-}=\epsilon_k+g_-n_0-g_{\uparrow\downarrow}\delta n\delta M_x$, $u_{k,\pm}=1$ and $v_{k,\pm}=0$.  In both cases, the condensate density $n_0=n-\delta n$, and $\epsilon_k=\hbar^2k^2/2m$. When $n_0$ vanishes, Eqs.~(\ref{self-consist-1}) and (\ref{self-consist-2}) become
\begin{align}
  n\lambda_{\rm T}^3 &= \zeta \big(\tfrac{3}{2} \big) + {\rm Li}_{3/2}\big(e^{\beta g_{\uparrow\downarrow}n \delta M_x}\big) \, , \label{nc_eqn1}
  \\
  n\lambda_{\rm T}^3 \delta M_x &= \zeta \big(\tfrac{3}{2} \big) - {\rm Li}_{3/2}\big(e^{\beta g_{\uparrow\downarrow}n \delta M_x}\big) \, , \label{nc_eqn2}
\end{align}
which recover Eqs. (\ref{density-eqn-0}) and (\ref{self-consist-0}) respectively at the BEC transition\;\footnote{For $|a_{\uparrow\downarrow}| \ll \lambda_{\rm T}$, the critical point of the BEC transition is very close to that of the ferromagnetic transition. By applying the expansion (\ref{expansion}) to the conditions of the phase transitions, we find that, at a given $T$, the leading difference between $n_{\rm C}$ and $n_{\rm M}$ is given by
\begin{align}
  n_{\rm C} - n_{\rm M} = - 8\pi \zeta\big(\tfrac{1}{2}\big) a_{\uparrow\downarrow}^2/\lambda_{\rm T}^5 \simeq 36.7 a_{\uparrow\downarrow}^2 /  \lambda_{\rm T}^5 \, .
\end{align} }.

In the representation of quasi-particles, we obtain the free energy density of the BEC phase
\begin{align}
 \mathcal{F} = \mathcal{F}_0 + \mathcal{F}_{\rm Q} - k_{\rm B}T  \sum_{\alpha=\pm} \int \frac{d^3k}{(2\pi)^3} \ln \left[ 1 + f(E_{k,\alpha}) \right] \, ,
\end{align}
with $\mathcal{F}_0 = \tfrac{1}{2}g_+n^2 + \tfrac{1}{4}g\delta n^2 + \frac{1}{4}g_{\uparrow\downarrow}\delta n^2 \delta M_x (2 - \delta M_x)$. $\mathcal{F}_{\rm Q}$ vanishes in the Hartree-Fock theory, but takes a nontrivial form in the Popov theory
\begin{align}
  \mathcal{F}_{\rm Q} =  \frac{1}{2}\int\frac{d^3k}{(2\pi)^3} \left[ E_{k,+} + E_{k,-} - 2\epsilon_k - g n_0 + g_{\uparrow\downarrow}\delta n\delta M_x + \frac{g^2n_0^2}{4\epsilon_k} + \frac{g_{\uparrow\downarrow}^2n_0^2}{4\epsilon_k} \right].
\end{align}
Physically, it originates from the quantum fluctuations associated with the zero-point motion of Bogoliubov  quasi-particles.
In the highly degenerate limit $n\lambda_{\rm T}^3\gg 1$, thermal atoms are negligible, accordingly,  $\mathcal{F}_0$ reduces to the mean-field energy density of the condensate $\mathcal{E}_0=\frac{1}{2}g_+n^2$, and $\mathcal{F}_{\rm Q}$ recovers the Lee-Huang-Yang correction to the ground state energy $\mathcal{E}_{\rm LHY} = 64 \left(g_+\sqrt{na_+^3} + g_-\sqrt{na_-^3}\right)n^2/15\sqrt{\pi}$. For the case of $|a_{\uparrow\downarrow}|$ comparable to $a$, the attractive and the repulsive mean-field energy mostly cancel out, and the Lee-Huang-Yang correction provides an important contribution to the equation-of-state.

Once free energy is obtained, chemical potential and pressure are readily evaluated according to the thermodynamic relations $\mu = \left(\partial \mathcal{F} / \partial n\right)_T$ and $P=\mu n - \mathcal{F}$. These quantities are used in the determination of the balanced conditions for the normal-superfluid phase separation [Eq.\;(8) of the main text].

\section{III.\quad M\lowercase{ulti}-V\lowercase{alued} B\lowercase{ehavior} N\lowercase{ear the} BEC T\lowercase{ransition}}
\begin{figure} [t]
\includegraphics{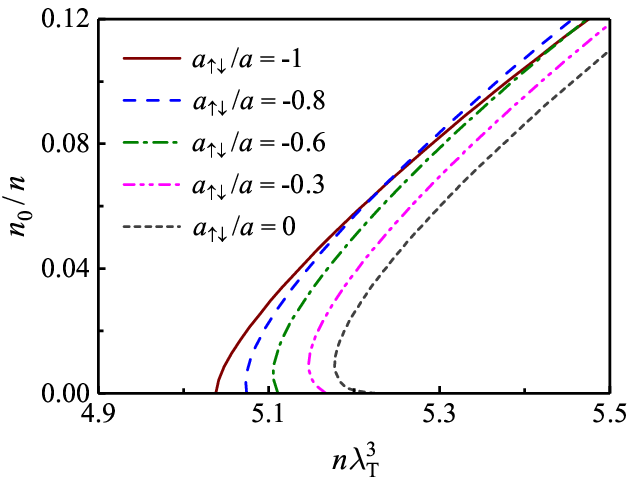}
\caption{Condensate fraction of spin-half bosons near the BEC transition for different values of $a_{\uparrow\downarrow}/a$. The calculations are based on the Popov theory. The results of the Hartree-Fock theory are similar. Parameters: $T=400$nK. Throughout this supplementary material,  the intra-spin scattering length $a$ and the atomic mass $m$ take the same values as in the main text. } \label{FigS1}
\end{figure}

It is well known that the self-consistency equations in the Popov theory and the Hartree-Fock theory usually have multiple solutions around the BEC transition, which results in a non-monotonic change in condensate fraction~\cite{ReentranceMF1,ReentranceMF2}. Such artifact also exists in our spin-half model.
However, as shown in Fig.\;\ref{FigS1}, the unphysical multi-valued region shrinks with increasing relative strength of inter-spin attraction. In particular, at $a_{\uparrow\downarrow}=-a$, the multi-valued behavior completely disappears in both the Popov theory and the Hartree-Fock theory\;\footnote{This result can be simply understood by noticing that, in the specific case of $a_{\uparrow\downarrow}=-a$, the lower branch of excitation spectrum $E_{k,+}$ is identical to that of the noninteracting system.}.
On the other hand, the density interval of spinodal decomposition enlarges as the ratio $|a_{\uparrow\downarrow}|/a$ increases [see Fig.\;1(a) in the main text]. These two opposite trends clearly show that the mechanical instability is a consequence of the inter-spin attractive interactions, rather than an artifact of the mean-field approximations.

\section{IV.\quad T\lowercase{hermodynamics of an} I\lowercase{ncoherent} B\lowercase{ose} M\lowercase{ixture}}

As mentioned in the main text, if the inter-spin conversion is prohibited, the transverse ferromagnetic order would not emerge spontaneously. Experimentally, the system prepared under this condition is an incoherent binary mixture, in which the atomic number of each component is separately conserved.

Now, let us briefly discuss the thermodynamic properties of such incoherent mixture. For simplicity, we consider atoms have the balanced population in each component and the interaction parameter $a+a_{\uparrow\downarrow}$ is not very close to zero\;\footnote{Here, we still use $\uparrow$ and $\downarrow$ to label the two components of the mixture and assume $a_{\uparrow\uparrow}=a_{\downarrow\downarrow}=a$.}.
In that case, the contribution of quantum fluctuations can be safely ignored, and the equation-of-state predicted by the Hartree-Fock theory is qualitatively reliable.

We start by assuming that the mixture is homogeneous.
Within the Hartree-Fock approximation, each component behaves like a single gas with the chemical potential having a mean-field shift due to the interspecies interactions. In the BEC phase, we have
\begin{align}
  \mu_\uparrow = \mu_\downarrow = \tfrac{1}{2}g\left( n +\delta n \right) + \tfrac{1}{2}g_{\uparrow\downarrow} n   \, , \qquad \text{(mixture)}
\end{align}
where $\delta n$ is the total density of thermal atoms given by
\begin{align}
  \delta n = \sum_\sigma \int \frac{ d^3 k}{(2\pi)^3}  f(E_{k,\sigma}) = 2\lambda_{\rm T}^{-3} \mathrm{Li}_{3/2}\big(e^{-gn_0/2k_{\rm B}T}\big)  \, , \qquad \text{(mixture)}
\end{align}
with $ E_{{\bf k},\uparrow} = E_{{\bf k},\downarrow} = \epsilon_k + gn_0/2$ being the excitation energy.  For comparison, in the transverse ferromagnetic BEC phase of spin-half bosons, the chemical potential reads,
\begin{align}
  \mu = \tfrac{1}{2} g\left(n + \delta n \right) + \tfrac{1}{2}g_{\uparrow\downarrow} \left( n + \delta n\delta M_x \right)   \, , \qquad \text{(spinor)}
\end{align}
where $\delta n$ and $\delta M_x$ are determined by Eqs.~(\ref{self-consist-1}) and (\ref{self-consist-2}). One can see that, in the spinor system,  the transverse magnetization of thermal atoms gives rise to an additional shift in chemical potential.

\begin{figure}[t]
\includegraphics{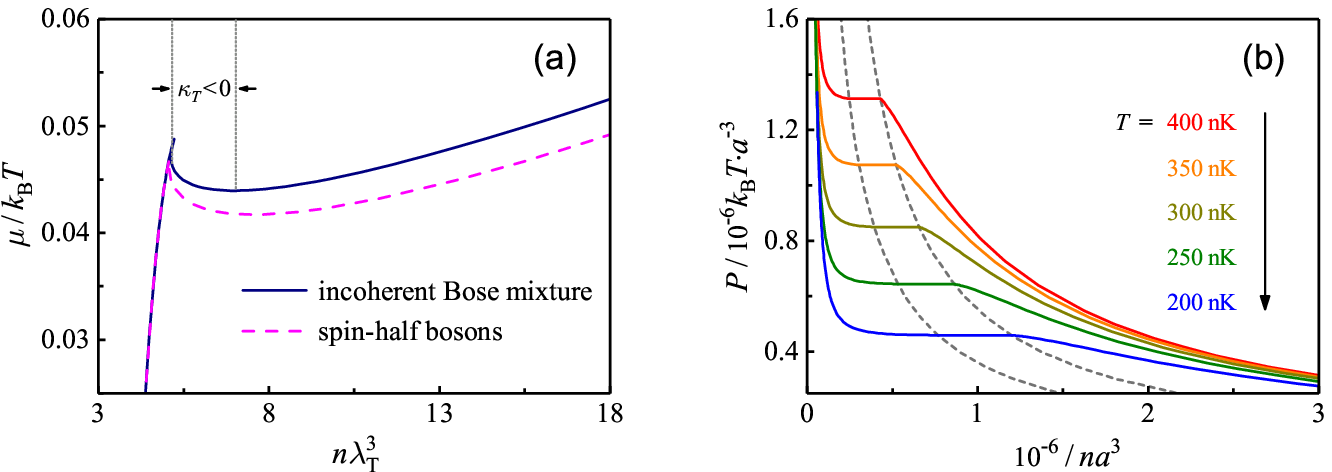}
\caption{(a) Isothermal $\mu$-$n$ curve of a homogeneous incoherent Bose mixture at $T=400$nK. For comparison, the results for spin-half bosons are also shown. (b) The isotherms with the normal-superfluid phase separation at different temperatures. The horizontal segments between the short-dashed lines correspond to the coexistence states. The calculations are based on the Hartree-Fock theory. Parameters: $a_{\uparrow\downarrow}=-0.8a=52a_0$. } \label{FigS2}
\end{figure}

In Fig.\;\ref{FigS2}(a), the isothermal $\mu$-$n$ curves of an incoherent Bose mixture and spin-half bosons are compared for the same interaction parameters. The two curves coincide in the non-degenerate regime but differ in the BEC phase\;\footnote{In the limit $n\lambda_{\rm T}^3\gg 1$, both curves approach to the same asymptotic result $\mu=g_+n$.}. Combine this numerical result with the thermodynamic relation $ \mathcal{F}(n)  = \int_{0}^{n} dn'\, \mu(n') $, one can deduce that (at given $n$ and $T$) the free energy of the incoherent BEC mixture is higher than that of the spin-half bosons. Such difference in equation-of-state can be understood by noticing that, in the spinor system, the inter-spin interactions produce an extra mean-field energy proportional to $\delta M_x^2$.

\begin{figure}[b]
\includegraphics{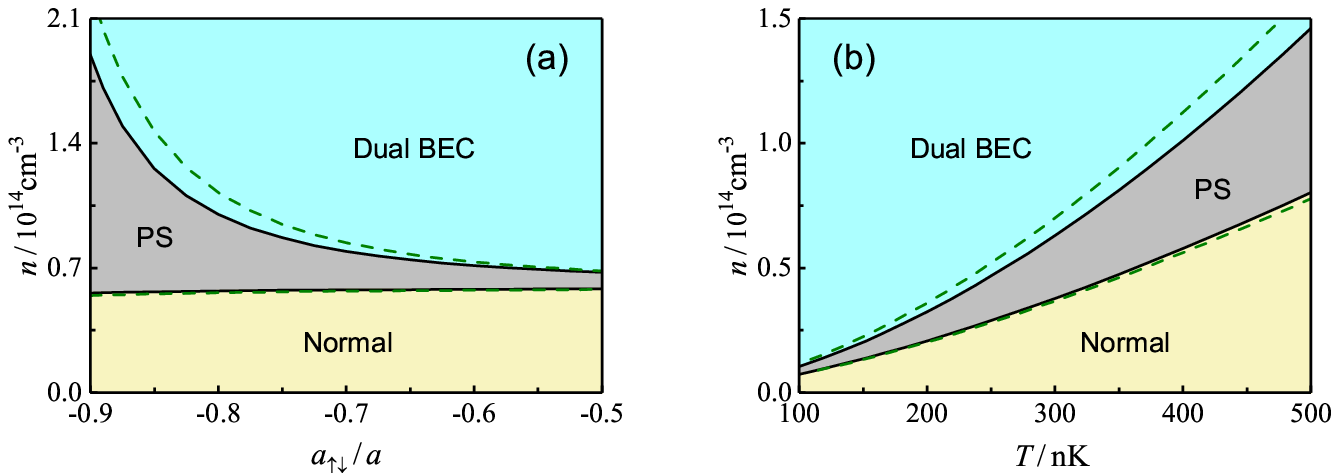}
\caption{Phase diagrams of a balanced incoherent Bose mixture for (a) fixed temperature at $T=400$nK, and (b) fixed interaction strength with $a_{\uparrow\downarrow}=-0.8a=52a_0$. The calculations are based on the Hartree-Fock theory. 
For comparison, the boundaries of phase separation region for spin-half bosons are also shown (dashed-lines).} \label{FigS3}
\end{figure}

The non-monotonic behavior of the $\mu$-$n$ curve indicates that the homogeneous state of the incoherent Bose mixture also suffers the spinodal decomposition within a medium range of density.
Therefore, the normal-superfluid phase separation occurs and results in the featured isotherms with horizontal portions [see Fig.~\ref{FigS2}(b)]. The corresponding phase diagrams are displayed in Fig.~\ref{FigS3}. Compared with the spin-half system, the area of coexistence region shrinks apparently for relatively stronger interspecies attraction and higher temperature. In the presence of  a harmonic trap, the phase separation can be experimentally detected in the similar way as discussed in the main text.

We finally remark that, if the interspecies conversion is practically allowed, the incoherent BEC mixture is only metastable in the sense that the formation of transverse magnetization would reduce the free energy. For bosons occupying two hyperfine sublevels of the same isotope, the transverse ferromagnetic BEC phase can be realized when an applied radio-frequency field driving the inter-spin transition is adiabatically switched off. Owing to the larger area of the coexistence region, it is more promising to observe the normal-superfluid phase separation in such a spinor configuration.



